\documentclass[onecolumn]{article}
\pdfoutput=1 

\usepackage{jinstpub} 

\usepackage{hyperref,longtable}

\usepackage{lineno}

\usepackage{url}
\usepackage{xcolor}
\usepackage{amsmath}
\usepackage{graphicx}
\usepackage{xfrac}
\usepackage{verbatimbox} 
\usepackage{float} 

\def\nuc#1#2{${}^{#1}$#2}

\title{Variable Energy X-ray Fluorescence Source}

\author[a,1]{S.R. Elliott,\note{Corresponding author.}}
\author[b]{E.M. Bond}
\author[c]{B. Dodson}
\author[b]{G. Rusev}
\author[a]{R. Massarczyk}
\author[d]{S.J. Meijer}
\author[a,2]{M. Stortini\note{Permanent Address: Physics Department, Yale University, New Haven, CT, 06511, USA}}
\author[c]{C. Wiseman}

\affiliation[a]{Physics Division, Los Alamos National Laboratory, Los Alamos, NM, 87545, USA}
\affiliation[b]{Chemistry Division, Los Alamos National Laboratory, Los Alamos, NM, 87545, USA}
\affiliation[c]{Center for Experimental Nuclear Physics and Astrophysics, and Department of Physics, University of Washington, Seattle, WA 98195, USA}
\affiliation[d]{Nuclear Engineering and Nonproliferation Division, Los Alamos National Laboratory, Los Alamos, NM, 87545 USA}

\emailAdd{elliotts@lanl.gov}

\abstract{
We detail the design of a variable energy, x-ray fluorescence source using a low activity (1.8$\times10^6$~dpm) \nuc{99}{Tc} $\beta$ source that irradiates thin foils. By rotating the source among foils of Ti, Zn, Nb, Ag, and Au, the device produces x rays between 4 and 70~keV at a rate near 1~Hz. When the source is placed in a storage position, the external radiation is non-detectable. The design of the shielding and rotation mechanism permits use in vacuum and at liquid nitrogen temperature. The design is intended for the study of the low energy response to radiation impinging upon Ge detector surfaces. The source will be useful for understanding the detector response in large-scale Ge arrays such as \textsc{Majorana} and LEGEND.
}


\begin{document}
\maketitle
\flushbottom

\section{Introduction}
\begin{figure}[H]
\centering
\includegraphics[width=0.8\textwidth]{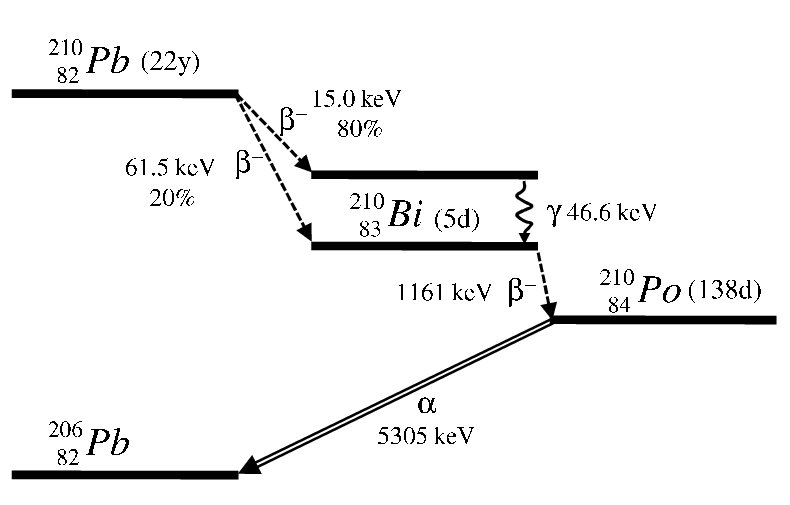}
\caption{$^{210}$Pb Decay Chain.}
\label{fig:LeadDecayChain}
\end{figure}

Low-background experiments searching for rare processes using Ge detectors (e.g. double beta decay)  have made significant contributions looking for new physics at low energy~\cite{Arnquist2022ExoticDM,Arnquist2022axion,Arnquist2022WFC,Chen2016,Li2013Texono,Zhang2022,Wang2022,Aalseth2012}, especially following the development of p-type, point-contact detectors~\cite{luk89,Barbeau2007,COOPER2011}. Models of the background contributions at energies below a few keV, however, have significant uncertainty. Data from the \textsc{Majorana Demonstrator}, for example, show that $^{210}$Pb is present as evidenced by the observation of its 46-keV $\gamma$ ray in the spectrum, as well as the observation of $^{210}$Po $\alpha$ emission at 5.3 MeV. Since $^{210}$Po is a daughter of $^{210}$Pb (Fig.~\ref{fig:LeadDecayChain}), and both, \nuc{210}{Po} and \nuc{210}{Bi}, have shorter half-lives, some fraction of the Po must be supported by the Pb decay. While the 46-keV $\gamma$ ray can penetrate the insensitive layer that encloses most of the detector surface to approximately 1~mm,  $\alpha$ particles cannot. However, the passivated layer surrounding the point contact, and separating the HV-biased n$^+$ surface, is only of the order of a $\mu$m thick permitting $\alpha$ observation. Since both radiations are observed, it is assumed that at least some of the $^{210}$Pb must irradiate the passivated surface. Although the spectrum includes the 46.5-keV $\gamma$ line from $^{210}$Pb, it does not observe the associated 10.8-keV x ray~\cite{Arnquist2022ExoticDM}. Furthermore, the 46-keV level decay transition has a large internal conversion coefficient leading to electron emission with energies near 30~keV, which are also not seen. These non-observations indicate an incomplete understanding of the response of the passivated surface to radiation, such as low energy $\gamma$s or electrons, that create charge near the surface. 

By scanning the surfaces of a Ge detector with x-rays of various energies, the depth-related effects of energy deposit as a function of position can be studied.  This motivates our design of a variable energy x-ray fluorescence source, with energies between $\sim$4-70 keV in a tunable device that can be operated in vacuum and at liquid nitrogen temperatures. Details on the design of this source are presented in this paper.

Although x-ray fluorescence sources have been commercially available in the past, we were unable to find one that met our present needs. As a consequence, we chose to design and build a custom source for this purpose. The literature is rich in discussion of x-ray fluorescence and we referred to a number of studies that guided us as we optimized the x-ray production. Reference~\cite{Valkovic1993} is an old summary the field of x-ray fluorescence and its use for analysis. The IAEA report~\cite{IAEA1970} discusses the relative fluorescence rates due to $\alpha$, $\beta$,  $\gamma$, and x-ray excitation. Meijer and Unger~\cite{Meier1976} estimated x-ray production from a number of sources and targets for fluorescence analysis. Reference~\cite{Manning1991} describes a 1-MeV electron beam to excite La x rays to image human coronary arteries and Refs.~\cite{Rieder1997,Buchner2003,Perrett2015} studied the performance of an x-ray fluorescence emission from a \nuc{244}{Cm} source exciting target material for the Mars Pathfinder program.

\section{Device Simulation}

Our design concept employs a $\beta$ source radiating carefully chosen foils to produce x-ray fluorescence. By assembling a wheel of differing material foils that can be spun relative to the radioactive source,  a variety of possible x-ray energies can be produced. We selected a group of 5 foils to provide high-rate $K_{\alpha}$ emissions somewhat evenly distributed between $\sim$4-70~keV. Our simulations indicated that a $\beta$ source would produce more x rays than an $\alpha$ or $\gamma$ source in agreement with Ref.~\cite{IAEA1970}

The ideal source for our needs is a pure $\beta$ decay to a stable daughter with a long half-life and an endpoint as low as possible, but high enough to ionize the Au K shell.
We considered two primary options, $^{14}$C (5730~yr, 156~keV) and $^{99}$Tc (2.1$\times10^{5}$~yr, 297~keV).
Our simulations indicated that $^{14}$C does not produce significant $K_{\alpha}$ emission in Au.
Therefore, we choose $^{99}$Tc  as our $\beta$ source.

The key design parameters include the radioactive source, its activity, the foil thicknesses, and the source-foil housing materials.
The figure of merit used to optimize the device was the predicted percentage of spectrum counts residing within the $K_{\alpha}$ peak for each foil.
The resulting energy spectra with their associated x-ray and background rates differ for each foil. Therefore, it was also necessary to design the device holding the source and foils to maximize the x-ray rate, while minimizing background. 
We simulated \nuc{99}{Tc} radiation of the foils to determine the thickness that maximized the $K_{\alpha}$ production rate. Prior to obtaining a \nuc{99}{Tc} sample, we compared the simulation to a measurement of an available \nuc{14}{C} $\beta$ source irradiating a Ti foil. The simulation result was within 20\% of the measured value indicating that the simulation worked well enough to guide our design choices. 

We simulated foils of Zn ($K_{\alpha}$ = 8.639~keV), Ag (22.163~keV),  and Au (68.804~keV) to establish the appropriate thickness, however, the final selection of foils included Ti and Nb to provide energies of interest. For the three foils simulated (Zn, Ag, Au), the optimum thicknesses were found to be 80~$\mu$m, 80~$\mu$m, and 250~$\mu$m, respectively. Readily available commercial foils come in pre-determined thicknesses and our final selections were chosen to be close to these simulation optima.

The simulation software used for this study was Geant4~\cite{Agostinelli2003}. Given that low-energy electromagnetic physics was the primary interest here, we used the physics list ``Shielding$\_$LIV''. A number of electromagnetic processes that were important to our studies are turned off by default in Geant4, and had to be enabled. It was also found that the Bearden data~\cite{Bearden1967} in Geant4 simulated the correct x-ray energies much better than the default fluorescence data.

\section{Device Design}
\begin{figure}[htb!!!]
\centering
\includegraphics[width=0.2\columnwidth,angle=90] {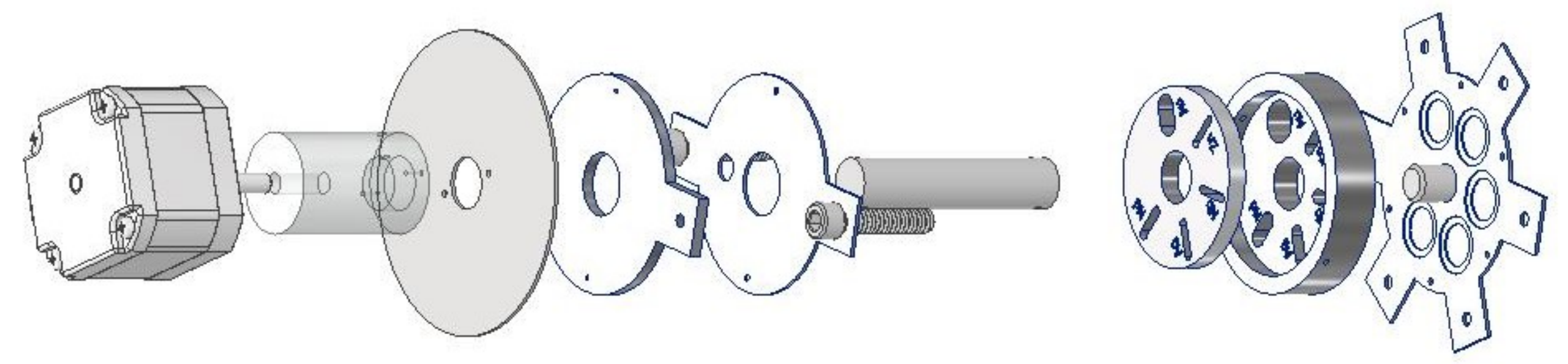}
\caption{\small{An exploded view of the device.}}
\label{fig:ExplodedView}
\end{figure}

\begin{figure}[htb!!!]
\centering
\includegraphics[width=0.45\columnwidth] {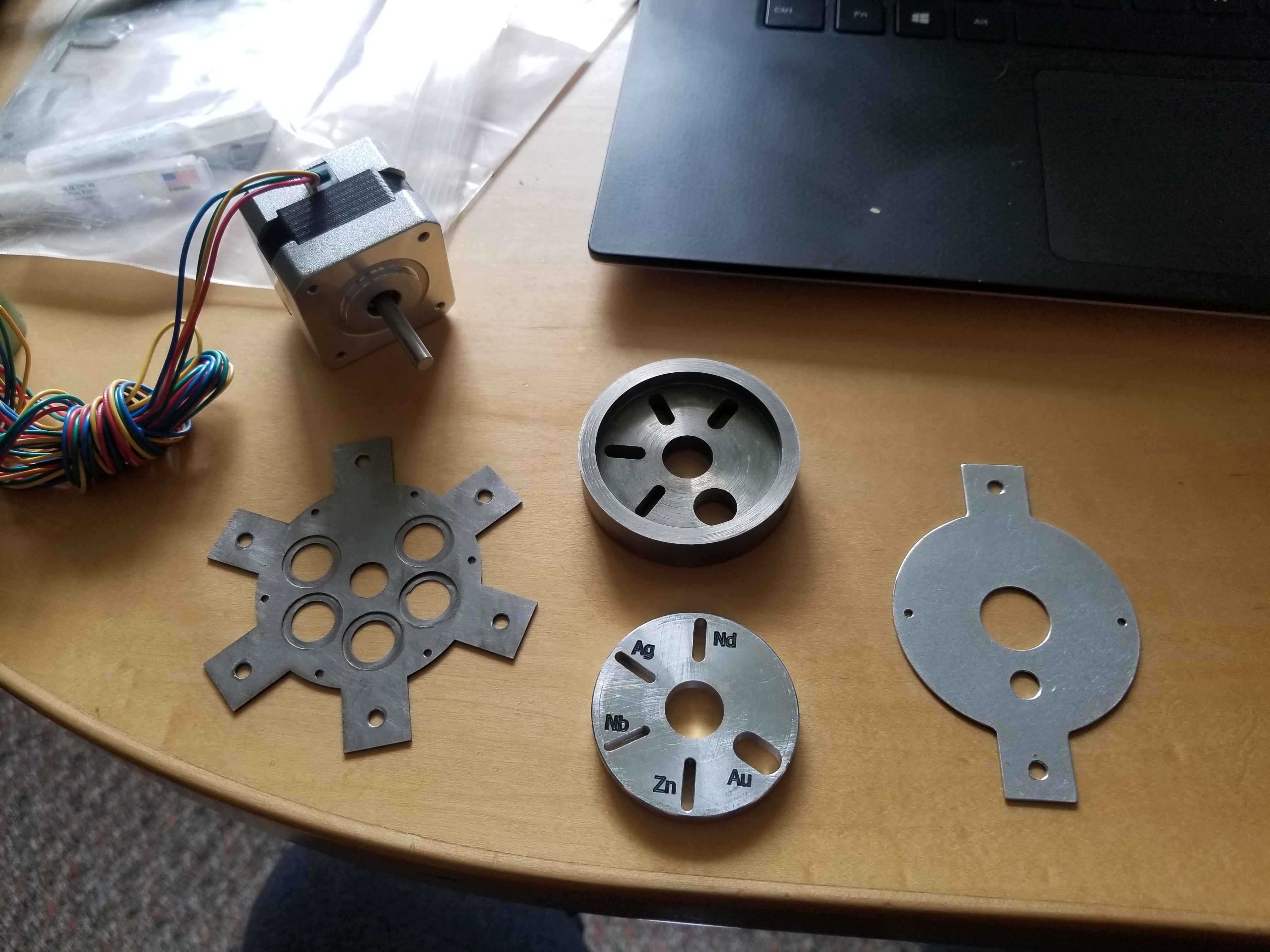}
\includegraphics[width=0.45\columnwidth] {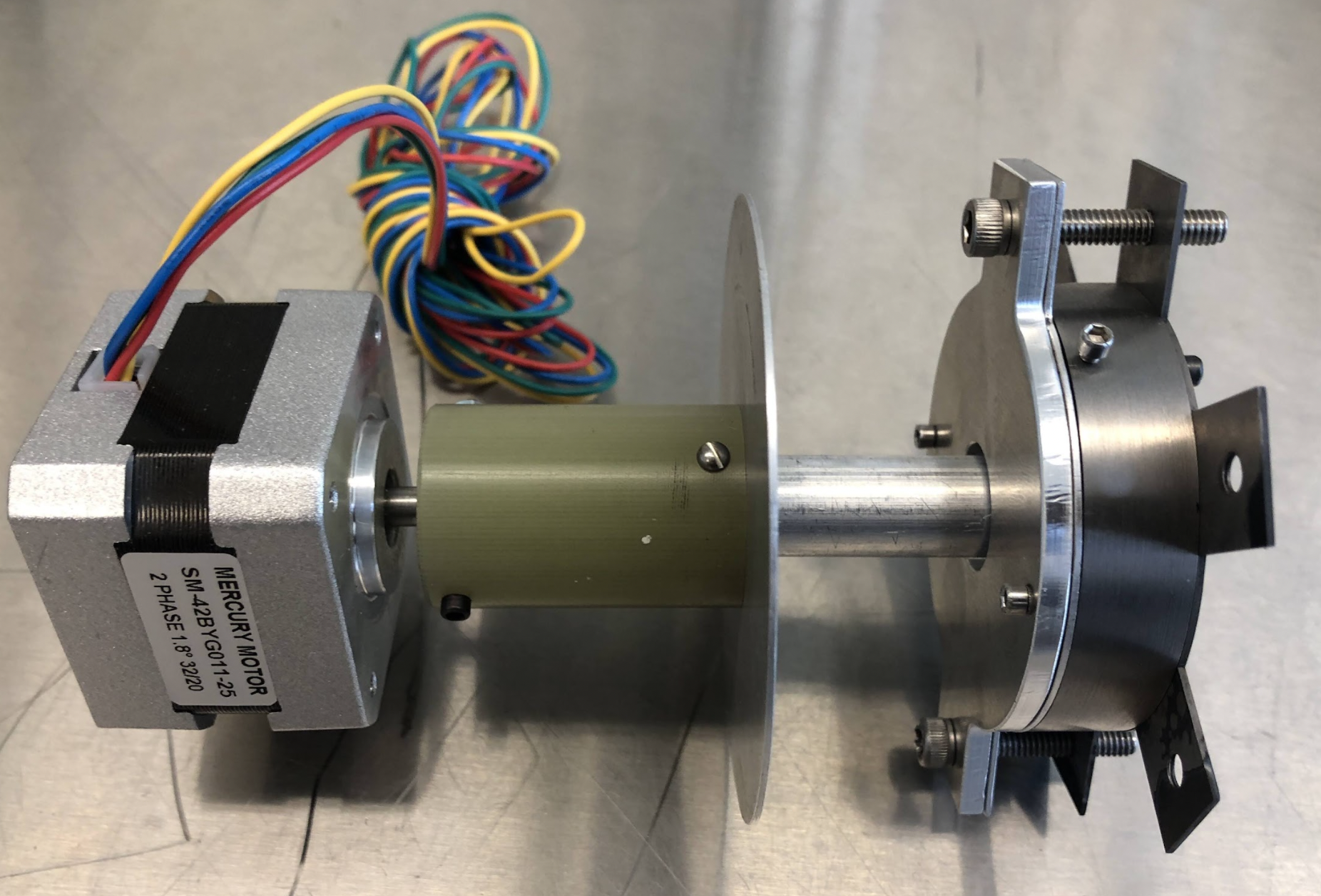}
\caption{\small{Left: The foil holder, W shielding disk, Al shielding disk and the 1~mm slice of the source sandwich. Right: The final assembled device.}}
\label{fig:Assembly}
\end{figure}

The construction materials for the device were chosen to minimize background arising from bremsstrahlung from the stopping $\beta$ rays.
The design is a graded-Z disk separating the source and foils. By surrounding the source with low Z material, bremsstrahlung is limited and $\beta$s that do not impinge on the foil are blocked. Surrounding this low-Z region, is a higher-Z material to shield any remaining bremsstrahlung. The design, shown in Figs.~\ref{fig:ExplodedView} and \ref{fig:Assembly} has an Al structure contained within W. Tungsten (4.5$\times10^{-6}$~m/(m~K)) and Al (21$\times10^{-6}$~m/(m~K)) have coefficients of linear expansion similar enough to address cryogenic engineering issues. Plastic could have been a good choice for the low-Z shield otherwise. The dimensions of the Al and W were determined by simulation to reduce background to less than 10\% of the signal rate in the lowest rate configuration with the Au foil. 

Starting from the x-ray emission end, the device consists of a W foil holder, an Al shielding disk, a W shielding disk, and a source holder consisting of two thin Al pieces. These parts are shown in Fig~\ref{fig:ExplodedView}. A drive shaft passes through the source holder to attach to the foil holder. When the shaft is turned the foil holder and shield disk rotate, while the source holder remains stationary. Therefore the choice of irradiated foil can change without altering the x-ray target spot.

The W foil holder has 5 through holes, 10.626~mm in diameter, centered on 0.5-mm-deep mounting wells of 14.5~mm diameter. The 10-mm diameter foils reside in these wells and the foil holder is bolted to the bottom of the W shielding disk trapping the foils next to the Al shielding disk. The Al shielding disk is 45.5~mm in diameter and 5~mm thick. It has 5 collimation slits for $\beta$ passage from the source to the foils. The W shielding disk is 55~mm in diameter but has a 46-mm diameter inner well to enclose the Al shielding disk. 
The overall thickness of the W shielding disk is 12~mm with the well depth being 5~mm to match the thickness of the Al shielding disk.  Slots for $\beta$ passage through the W shielding disk match those in the Al shielding disk. The slots have varying dimensions chosen to provide similar x-ray emission rates. The source holder consists of two Al plates, one 3~mm thick and one 1~mm thick, which form a sandwich trapping the source. The thicker piece has a 1.1-mm deep well feature, 10.09~mm in diameter, to house the thin titanium foil with the \nuc{99}{Tc} deposit (not to be confused with the Ti target foil). The 1~mm piece has a through hole, 6.5~mm diameter, that allows emitted $\beta$s to enter the shielding slots. Tabs on the source holder and foil holder permit a secure arrangement, when desired. A fully shielded sixth position places the \nuc{99}{Tc} in a safe position with no detectable emanated radiation.

Figure~\ref{fig:ExplodedView} shows an exploded view of the device. The photos in Fig.~\ref{fig:Assembly} show the source parts and the assembled device attached to a stepper motor via the shaft. Along the shaft is an infrared shield to minimize radiative heating of the device when it is in use near a Ge detector. Also shown is a G-10 shaft adapter to minimize heat conductance down the shaft.

\section{Source Preparation}
The \nuc{99}{Tc} electrodeposition method was developed based on a procedure by Voltz~\cite{Voltz1967}, using an electrodeposition cell that has been previously described~\cite{Fitzpatrick2008}.   A 2.77-mL aliquot of the \nuc{99}{Tc} solution (obtained from Eckert \& Ziegler~\cite{Eckert}; 2.22$\times10^7$~dpm in 5~g, 0.001~M NaOH) was dried to $\sim$0.2~mL and transferred to the electrodeposition cell, followed by 1.35~mL of the electrodeposition solution (2~M (NH$_4$)$_2$SO$_4$, adjusted to pH = 1).  The electrodeposition was carried out for 60~minutes at 0.1~A using a 51~$\mu$m Ti foil as the cathode.  The resulting black deposit was dried on a hotplate set to 110~C and assayed by beta analysis: 1.77$\times10^6$~dpm for a 15\% yield.  To reduce the risk of contamination, the deposit was covered by a transparent 2~$\mu$m PET film with an 8~$\mu$m thick acrylic adhesive from PolyK Technologies LLC.

\section{Results and Measurements}
We measured the x rays from the source with a FAST-SDD\textsuperscript{\textregistered} silicon drift detector from Amptek Inc.~\cite{Amptek}. The detector has a thickness of 0.5~mm and collimated to an area of 50~mm$^2$. It is placed in a vacuum in a casing with a 0.5-mil-thin beryllium window and cooled at 220~K. The signals from the detector were processed by a stand-alone digital pulse processor and the Apmtek data acquisition program.
We measured the efficiency of the detector with a set of Eckert \& Ziegler standard sources of \nuc{54}{Mn}, \nuc{57}{Co}, \nuc{75}{Se}, \nuc{109}{Cd}, \nuc{133}{Ba}, \nuc{139}{Ce}, \nuc{153}{Gd}, and \nuc{241}{Am}.

\begin{figure}[H]
  \centering
  \includegraphics[width=0.45\columnwidth]{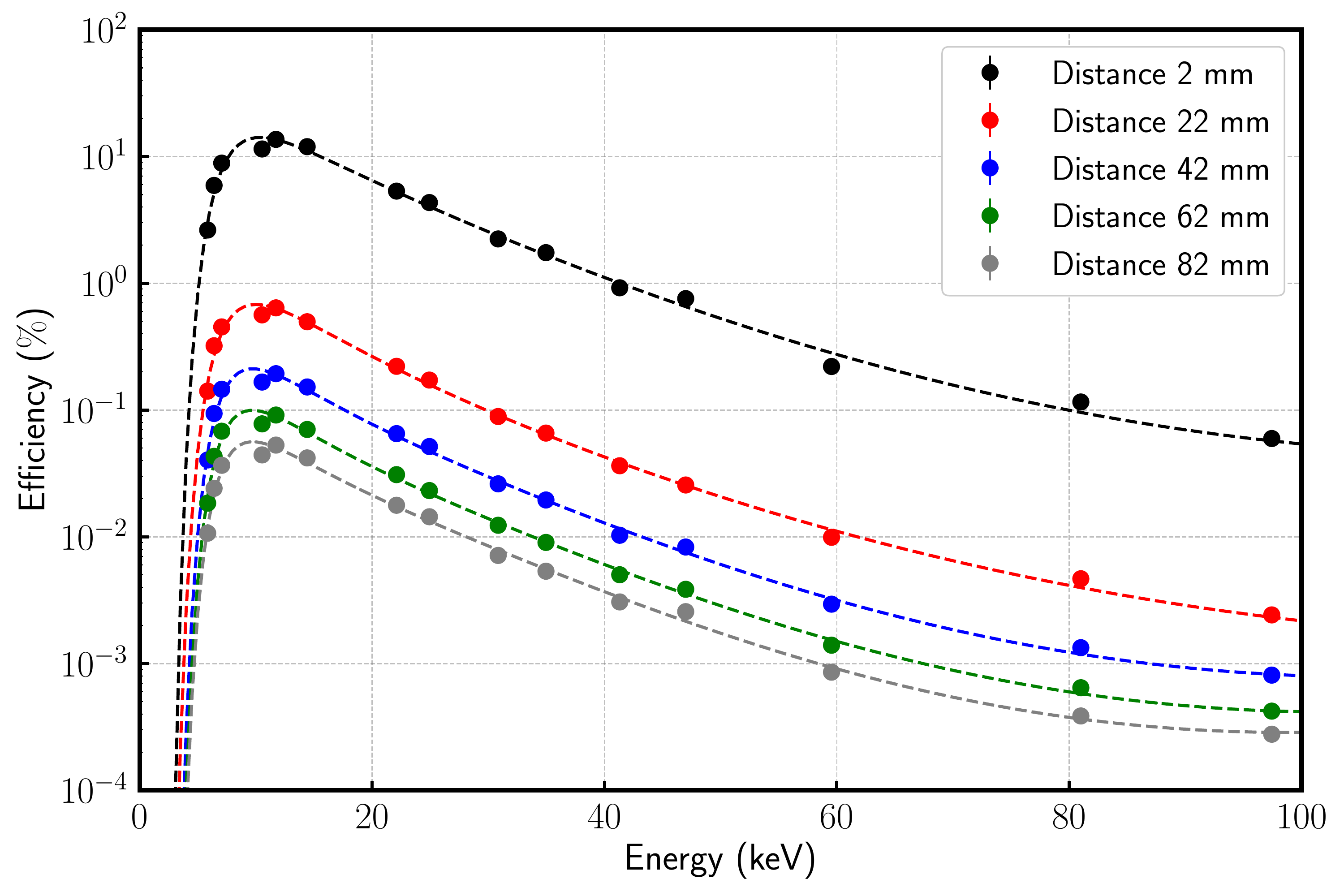}
  \includegraphics[width=0.45\columnwidth]{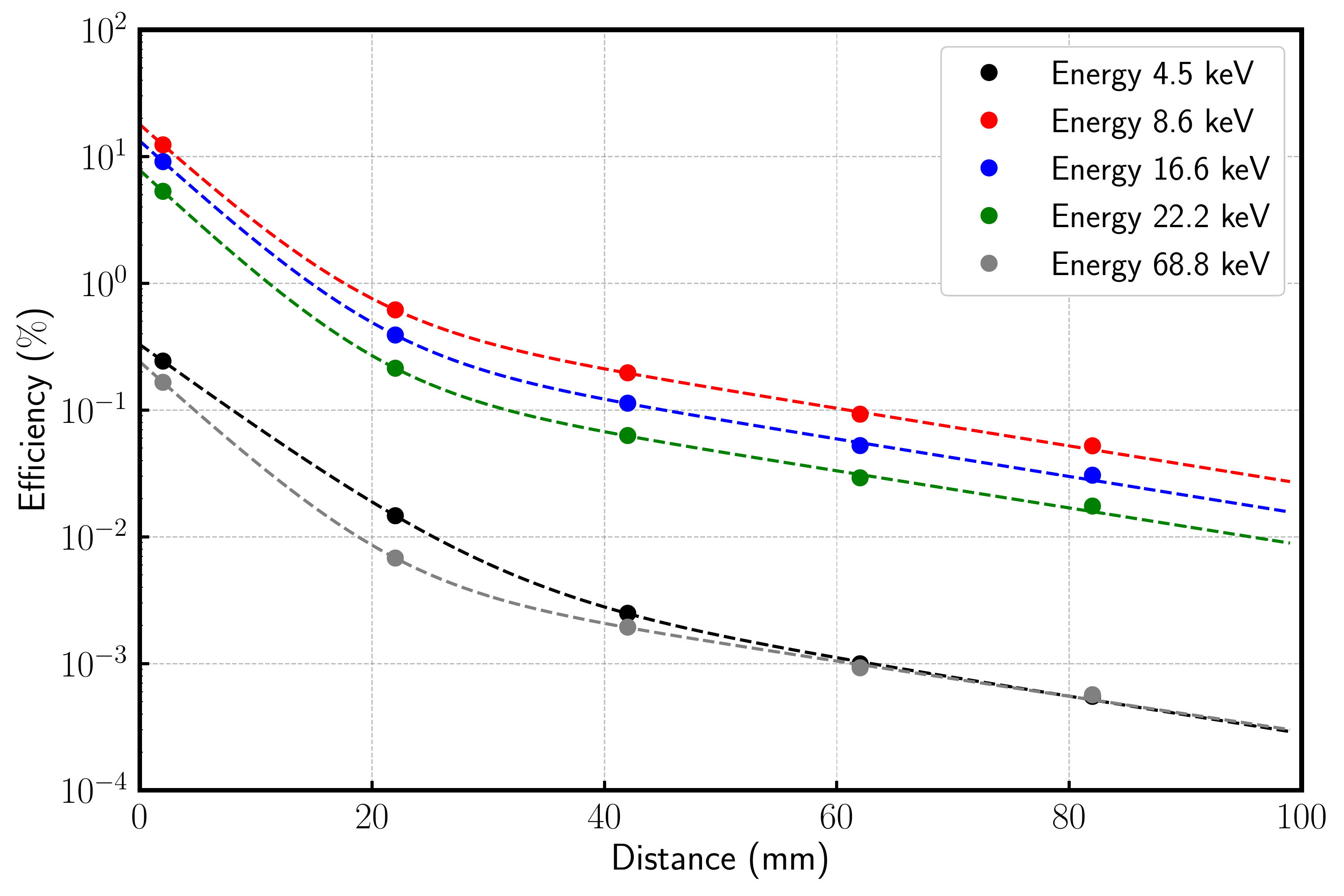}
  \caption{Left: the efficiency curves for the detector as a function of energy for 5 separation distances. Right: The efficiency for our 5 lines of interest as a function of distance. }
  \label{fig:Efficiency}
\end{figure}

The efficiency ($\epsilon$) for detecting x rays with this detector as a function of energy ($E$) and distance ($D$) is shown in Fig.~\ref{fig:Efficiency}. The curves through the data points are an empirical parameterization to aid in interpolating between data points. The results of this interpolation are given in Table~\ref{tab:Rates} where the quoted uncertainty includes both contributions. The parametric formulas are,
\begin{equation}
\epsilon(E) = \sum_{k=0}^{5}a_k \cdot \mathrm{log(E)}^k  
\label{eqn:EnergyParameters}
\end{equation}
and 
\begin{equation}
\epsilon(D) = \delta \cdot e^{\eta D} + \kappa \cdot e^{\lambda D}
\label{eqn:DistanceParameters}
\end{equation}
\noindent with the parameter fit values given in the Appendix.	
The measured points in the efficiency curves are known rather well (0.5\% or better). The efficiency uncertainty for our data, however, is dominated by the uncertainty in the placement of the source relative to the detector as the geometry for our device was not optimum for this purpose. Our placement estimate is 2$\pm$1~mm for the Ti and Zn foils, and 5$\pm$1~mm for the others. We used the position dependent parameterization to propagate this position uncertainty to estimate the efficiency uncertainty.

\begin{figure}[H]
  \centering
  \includegraphics[width=0.45\columnwidth]{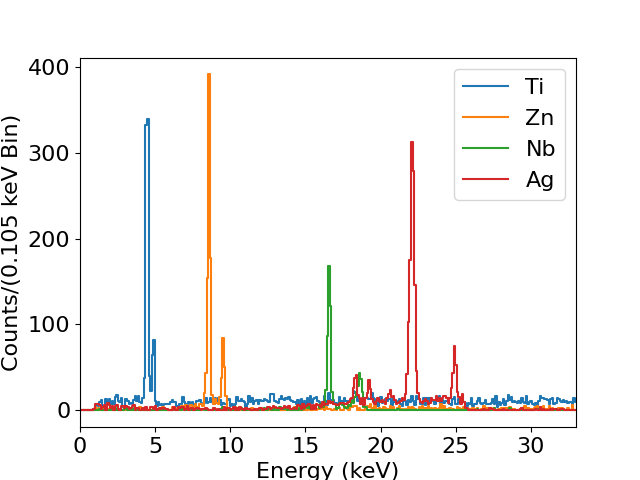}
  \includegraphics[width=0.45\columnwidth]{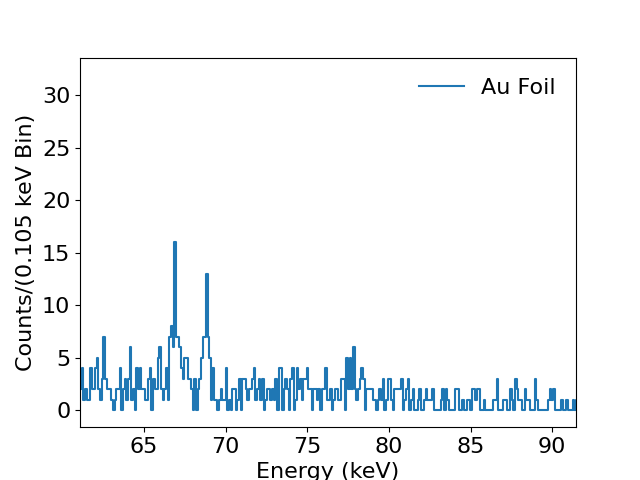}
  \caption{Example spectra from Ti, Zn, Nb and Ag (left), where the $K_{\alpha}$ and $K_{\beta}$ lines are visible and Au (right), where K$_{\alpha_2}$, K$_{\alpha_1}$ and K$_{\beta_1}$ are seen. }
  \label{fig:Spectra}
\end{figure}

Figure~\ref{fig:Spectra} shows example spectra. All spectra show signs of the $K_{\alpha}$ and $K_{\beta}$ lines, though the Au result is weak due to the low efficiency and reduced production rate of the higher energy x ray. Table~\ref{tab:Rates} summarizes the measured results with a listing of the x-ray production per $\beta$ decay of the \nuc{99}{Tc}. The detector efficiency is not precise at the low energy Ti x ray, as there were no efficiency calibration points used below 5.9~keV, the efficiency here was extrapolated, rather than interpolating. The detection efficiency also changes rapidly in this region, together resulting in lower confidence in the efficiency value at this energy.

\begin{table*}[htp]
\caption{The measurement parameters and results. See text for discussion of the Ti results and concerns regarding efficiency. The line energies are from Ref.~\cite{Deslattes2022} where we use the experimental values.}
\begin{center}
\begin{tabular}{lcccccccc}
\hline
Foil 	&	Thick. 		& K$_{\alpha}$ Ener.		& K$_{\beta}$ Energy	&	Live 				& Counts 			& Efficiency						& x ray  & Rate/\nuc{99}{Tc} Decay	\\
	&	($\mu m$)		& (keV)				& (keV)				&	Time (s)			&				&	(\%)						&Rate (Hz)	&	$\times10^{-5}$ (Hz)		\\
\hline\hline
Ti	&	\phantom050 	& \phantom04.511		& \phantom04.932		& \phantom0\phantom0571	&707$\pm$27		& 0.16$\pm$0.02				& 800$\pm$100	& (2600$\pm300)$	\\
Zn	&	\phantom075 	& \phantom08.639		&\phantom09.572		& \phantom05979			& 738$\pm$27		&7.16$^{+1.41}_{-1.17}$			& 1.72$^{+0.35}_{-0.29}$	& (5.84$^{+1.17}_{-0.98})$	\\
Nb	&	100 			& 16.615				& 18.622				& \phantom03602			& 411$\pm$20		&9.11$^{+1.88}_{-1.55}$			&1.25$^{+0.27}_{-0.22}$	& (4.25$^{+0.90}_{-0.75})$	\\
Ag	&	\phantom075 	& 22.163				& 24.942				& \phantom08088			& 986$\pm$31		&5.31$^{+1.13}_{-0.92}$			&2.30$^{+0.49}_{-0.40}$	& (7.78$^{+1.67}_{-1.37})$	\\
Au	&	259 			& 68.804				&77.980				&49880	& \phantom046$\pm$7\phantom0		&0.17$\pm0.03$				&0.54$\pm0.12$	& (1.84$\pm0.42)$	\\
\hline\hline
\end{tabular}
\end{center}
\label{tab:Rates}
\end{table*}%

\section{Summary}
We have constructed and tested a variable x-ray source using $\beta$ ray fluorescence producing an x-ray emission rate of about 1~Hz between 4 and 70~keV. A low radioactivity \nuc{99}{Tc} deposit provided the $\beta$ rays and when this activity is placed in its storage position, the external radiation is non-detectable. The device was designed to be used in vacuum at 80~K and is intended to be used to explore the  low energy response of a Ge detector passivated surface to radiation.

\section*{Acknowledgements}
We thank Tom Burritt for his machining expertise. We gratefully acknowledge the support of the U.S.~Department of Energy through the LANL Laboratory Directed Research and Development (LDRD) Programs for this work. This material is based upon work supported by the U.S.~Department of Energy, Office of Science, Office of Nuclear Physics under Federal Prime Agreement LANLEM78, and under award number DE-FG02-97ER41020.

\section*{Appendix: Efficiency Parameters}
\label{app:EfficiencyParameters}

Table~\ref{tab:EnergyParameters} gives the values for the various constants for the equation for $\epsilon(E)$ and Table~\ref{tab:DistanceParameters} gives the constants for $\epsilon(D)$.

\begin{table}[htp]
\caption{The parameters for $\epsilon(E)$ resulting from the fits with Eq.~\ref{eqn:EnergyParameters}.}
\begin{center}
\begin{tabular}{lccccc}
\hline
 	&2 mm 	& 22 mm	& 42 mm 	& 62 mm 	& 82 mm			\\
\hline\hline
a$_0$	&-97.07	&-93.27	&-131.6	&-140.6	&-144.9	\\
a$_1$	&151.8	&139.4	&204.3	&218.4	&226.0	\\
a$_2$	&-91.4	&-82.35	&-126.5	&-135.9	&-141.8	\\
a$_3$	&27.45	&24.16	&38.81	&41.89	&44.10	\\
a$_4$	&-4.14	&-3.563	&-5.939	&-6.435	&-6.832	\\
a$_5$	&0.2485	&0.2096	&0.3606	&0.3919	&0.4195	\\
\hline\hline
\end{tabular}
\end{center}
\label{tab:EnergyParameters}
\end{table}%

\begin{table}[htp]
\caption{The parameters for $\epsilon(D)$ resulting from the fits with Eq.~\ref{eqn:DistanceParameters}.}
\begin{center}
\begin{tabular}{lccccc}
\hline
 				&	Ti 		& Zn		&	Nb 	& Ag 	& Au			\\
\hline\hline
$\delta$			&0.008194	&17.11	&12.83	&7.562	&0.00710	\\
$\eta$			&-0.3373		&-0.194	&-0.195	&-0.199	&-0.0319	\\
$\kappa$			&0.3108		&0.799	&0.4543	&0.249	&0.234	\\
$\lambda$			&-0.1543		&-0.0341	&-0.0340	&-0.0336	&-0.1938	\\
\hline\hline		
\end{tabular}
\end{center}
\label{tab:DistanceParameters}
\end{table}%

\bibliographystyle{elsarticle-num}

\bibliography{XRFsource.bib}

\end{document}